\documentclass[twocolumn,superscriptaddress,aps,showpacs,floatfix,prl,12pt]{revtex4-1}
\usepackage[T1]{fontenc}
\usepackage{amssymb,amsmath}
\usepackage{graphicx}
\usepackage{color}
\usepackage{bbm}
\usepackage{bbold}
\usepackage{soul}
\usepackage[dvipsnames]{xcolor}
\usepackage{ulem}
\newcommand{\new}[1]{\textcolor{black}{#1}}

\newcommand{\expect}[1]{\langle #1 \rangle}
\newcommand{\ket}[1]{| #1 \rangle}
\newcommand{\bra}[1]{ \langle #1 |}

\begin{document}
\title{Reconstructing quantum states with quantum reservoir networks}

\author{Sanjib Ghosh}
\email{sanjibghosh87@u.nus.edu}
\affiliation{School of Physical and Mathematical Sciences, Nanyang Technological University 637371, Singapore}
\author{Andrzej Opala}
\affiliation{Institute of Physics, Polish Academy of Sciences, Al. Lotnik{\'o}w 32/46, PL-02-668 Warsaw, Poland}
\author{Micha{\l} Matuszewski}
\affiliation{Institute of Physics, Polish Academy of Sciences, Al. Lotnik{\'o}w 32/46, PL-02-668 Warsaw, Poland}
\author{Tomasz Paterek}
\affiliation{Institute of Theoretical Physics and Astrophysics, Faculty of Mathematics, Physics and Informatics, University of Gda\'{n}sk, 80-308 Gda\'{n}sk, Poland}
\author{Timothy C. H. Liew}
\email{timothyliew@ntu.edu.sg}
\affiliation{School of Physical and Mathematical Sciences, Nanyang Technological University 637371, Singapore}

\begin{abstract}
Reconstructing quantum states is an important task for various emerging quantum technologies. The process of reconstructing the density matrix of a quantum state is known as quantum state tomography. Conventionally, tomography of arbitrary quantum states is challenging as the paradigm of efficient protocols has remained in applying specific techniques for different types of quantum states. Here we introduce a quantum state tomography platform based on the framework of reservoir computing. It forms a quantum neural network, and operates as a comprehensive device for reconstructing an arbitrary quantum state (finite dimensional or continuous variable). This is achieved with only measuring the average occupation numbers in a single physical setup, without the need of any knowledge of optimum measurement basis or correlation measurements. 
\end{abstract}

\maketitle

\section{Introduction}
The interconnectivity of nodes in neural networks allows them to represent data in a high dimensional effective or feature space, in which they can learn to perform complicated transformations based on examples of desired output from given input. \new{Typically the signal at each node of a neural network is encoded in some analogue variable. In a physical implementation, this could represent the action potential of a biological network, the charge of a transistor in an electronic system, or the amplitude of light in an optical network.}

In systems where the physical size of a network node is small, comparable to the de Broglie wavelength, the laws of quantum physics come into play. The state of each node is no longer characterized by a single analogue variable, but by a quantum state residing in the Hilbert space. Effectively, the Hilbert space allows the traditional feature space of a neural network to become exponentially larger. Consequently, quantum neural networks~\cite{Biamonte17} show speed-up of both learning~\cite{Dunjko16,Paparo14}  and (theoretically) problem solving efficiency~\cite{Neigovzen09,Benedetti16,Alvarez-Rodriguez17,Fujii17}. However, as the systems for building such networks typically require nanoscale precision to engineer correctly the couplings between network nodes, they are not easy to come by. If such controllability can be achieved, then quantum neural networks are powerful architectures, which can enhance quantum computers~\cite{Farhi18,Grant18} and quantum annealers~\cite{Amin18}. Yet, since there is only limited access to even small-scale quantum computers, many applications of quantum neural networks are unexplored. The majority of works focus on how quantum systems can be used for classical tasks, however, they could also be applicable to tasks in quantum information processing, i.e., tasks that are not only enhanced by the availability of the Hilbert space but cannot be performed without the Hilbert space to begin with.

Among the forms of classical recurrent neural networks, reservoir computing emerged as particularly suitable for hardware implementations in a wide variety of systems~\cite{Tanaka19}. The training of recurrent neural networks is often computationally inefficient and can be a non-converging process~\cite{Lukosevicius12,Bengio94}. In contrast, reservoir computing uses a randomly connected reservoir as a processing unit where the input signal is to be fed. Training takes place only at the readout level and keeps the reservoir itself unchanged. Consequently, training of a classical reservoir computer is relatively straightforward and computationally efficient~\cite{Lukosevicius12}. The concept was recently generalized to quantum systems and shown to allow the classification of quantum states as entangled or separable~\cite{Ghosh19} \new{(entanglement is a property of quantum states empowering communication and computation)}. While recognizing entanglement is an important task in quantum information, it is far from a complete characterization of a quantum state. \new{While in classical physics a state is determined by measuring some set of characteristic quantities (e.g., the number of particles or intensity in each mode) that should be the same if the same measurement is performed on identical states, in quantum physics a state does not have well defined characteristics before measurement and measurements performed on identical states can have different results. A quantum state is thus defined by a distribution of different possible measurement outcomes and the possible correlations between those measurements. The most general quantum state is typically represented by a density matrix and the process of reconstructing this density matrix from multiple copies of a quantum state is known as quantum state tomography. Whether with finite-dimensional or continuous variable systems, quantum state tomography requires the processing of outcomes (complex data) of measurements in a complete set of bases~\cite{Shadbolt11,Lvovsky09,James01,Haffner05,Lu07}.} The number of required measurement bases increases exponentially with the increasing \new{size of a quantum system (e.g., number of qubits or number of quantum modes)}. Noting that performing measurements in different bases amounts to reconfiguring an experimental set-up, quantum state tomography becomes exceedingly challenging as the number of measurement bases grows. While adaptive and self-guided methods~\cite{Ferrie14,Chapman16,Qi17} and neural network protocols used on classical data obtained from independent experiments~\cite{Torlai18,Quek18,Carrasquilla19} or methods taking advantage of specific properties of finite-dimensional density matrices~\cite{Gross10,Cramer10} reduce the number of required measurement bases, they still need many measurements in different bases to fully reconstruct quantum states. Protocols based on a many-outcome measurement do avoid multiple reconfigurable measurements~\cite{Oren17,Titchener18}, however, complexity remains in single photon detections, their correlation measurements, and in the reconstruction method that depends on the specific  transformation used in the experiment. Advanced tomographic schemes in the continuous variable domain, e.g., schemes based on measurements on displaced quantum states~\cite{Hofheinz09} or via a Lagrange interpolation method~\cite{Landon-Cardinal18}, also require measurements in different bases in phase space. Moreover, none of these schemes are universally applicable for states representing both finite-dimensional and continuous-variable systems.

\begin{figure}[h]
\centering
\includegraphics[width=1\columnwidth]{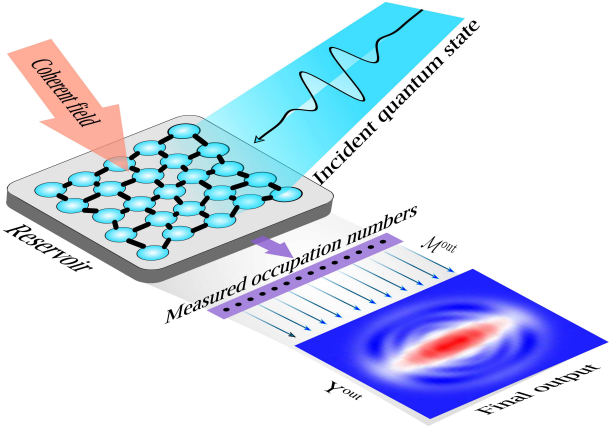}
\caption{The scheme of quantum reservoir state tomography. Here the reservoir is a fermionic lattice with random inter-site couplings. The input quantum state is taken in the form of an optical field incident on the reservoir. The incident field is coupled to the reservoir in cascade. The reservoir is excited with an additional resonant pump. The occupation numbers of the fermionic sites provide the readout elements, which then yield the final output $\textbf{Y}^\text{out}$ in the form of reconstructed density matrices or Wigner functions. The output weight matrix $\mathcal{M}^\text{out}$ is to be obtained through training.}
\label{TomoScheme}
\end{figure}

Here we present quantum reservoir state tomography (QRST) as a platform for universal quantum state reconstruction.
We consider a device that receives quantum information in the form of an optical field, which merges into a quantum reservoir network. The readout elements of the device are then provided by the occupation numbers measured on the reservoir (see the scheme in Fig.~\ref{TomoScheme}) rather than requiring any correlation measurements. This scheme constitutes a quantum version of reservoir computing that performs a quantum task, but here without assuming the pre-existence of a quantum computer. For QRST, while the quantum features of a reservoir allow to simplify the experimental protocol to a single measurement process, its reservoir computing framework enables to simplify the reconstruction subsequent to the measurement and guarantees its universal success for any quantum states in finite-dimensional and continuous-variable domains.

\section{Results}
Our scheme of quantum tomography using a quantum reservoir is schematically described in Fig.~\ref{TomoScheme}, where an input quantum state, represented by the density matrix $\rho_\text{in}$, is incident on a quantum network excited with a uniform coherent field $P$. We consider a 2D lattice of quantum dots (two level systems) for the quantum network (reservoir), represented by the Hamiltonian: 
\begin{eqnarray}
\hat{H} =  \sum_{\langle ij \rangle} J_{ij} \left(  \hat{c}_i^\dagger \hat{c}_j + \hat{c}_j^\dagger \hat{c}_i  \right)  + P\sum_i \left( \hat{c}_i^\dagger  + \hat{c}_i   \right)  
\end{eqnarray}
where $\hat{c}_i$ is the fermionic field operator (represents the quantum dots) at site $i$ and $J_{ij}$ are the hopping amplitudes between the nearest neighbour sites $i$ and $j$. We consider that $J_{ij}$ are randomly distributed between positive and negative values such that the spectral radius (largest absolute eigenvalue) of the hopping part of the Hamiltonian is $\tilde{J}$. The dynamics of the system can be described by the quantum master equation:
\begin{eqnarray}
i\hbar \dot{\rho} &=& [\hat{H},\rho] + \frac{i\gamma}{2} \sum_j \mathcal{L}(\hat{c}_j)   + i\Theta(t-t_1) \sum_k  \frac{\eta_k} {2\gamma} \mathcal{L}(\hat{a}_k) \nonumber \\
&+& i  \Theta(t-t_1) \, \sum_{k,j} \mathcal{M}^\text{in}_{jk} \left( [\hat{a}_k\rho, \hat{c}_j^\dagger ] +  [ \hat{c}_j, \rho \hat{a}_k^\dagger] \right) 
\label{MasterEq}
\end{eqnarray}
where  $\hat{a}_k$ are the field operators of the input modes (which can be bosons or fermions), $\rho$ is the combined density matrix representing the reservoir and the input modes, and $ \mathcal{L}(\hat{x}) = 2\hat{x}\rho\hat{x}^\dagger - \hat{x}^\dagger \hat{x}\rho -  \rho\hat{x}^\dagger \hat{x}$ is the Lindblad superoperator for a field operator $\hat{x}$. On the right-side of Eq.~\ref{MasterEq}, the first term represents the coherent Hamiltonian evolution of the reservoir, the second term represents the decay in the reservoir modes with the rate $\gamma/\hbar$, the third term is representing decay in the input modes with rates $\eta_k/(\hbar\gamma)$ due to the cascaded coupling between the input modes and the reservoir, represented by the remaining terms~\cite{Carmichael93,Gardiner93}. The parameters $\eta_k = \sum_j (\mathcal{M}^\text{in}_{jk} )^2$ are set by the cascaded formalism to ensure that the emitted photons from the input modes are only absorbed by the reservoir. $\mathcal{M}^\text{in}_{jk}$ are the input weights randomly chosen from the interval $[0,\omega]$. $\Theta(t-t_1)$ is the Heaviside function signifying that the cascaded coupling between the input modes and the reservoir fermions starts at $t=t_1$, where $t_1$ is an initial time interval. 

We perform QRST in four steps: (1) initially, for $0\le t < t_1$, the reservoir is only excited with the uniform field $P$, such that the reservoir at time $t_1$ reaches a steady state. (2) Then we activate the cascaded coupling between the input modes and the reservoir through the Heaviside function. This coupling moves the reservoir out of its initial steady state. (3) In this transient, we measure the expectation values of the occupation numbers $n_j=\langle \hat{c}^\dagger_j \hat{c}_j\rangle $ at time $t=t_1+\tau$ for all fermions in the reservoir. These occupation numbers provide the readout elements for the processing. (4) We evaluate the desired output $\textbf{Y}^\text{out} = \mathcal{M}^\text{out} \vec{n} + \vec{m}$ from the readout elements, where the output weight matrix ${\mathcal{M}}^\text{out}$ and state-independent constant vector $\vec{m}$ is to be determined through training.

We first present results of simulations of this tomographic scheme in various situations to demonstrate its universality and then explain the observed features with a mathematical proof of the process.


\begin{figure*}[]
\includegraphics[width=2\columnwidth]{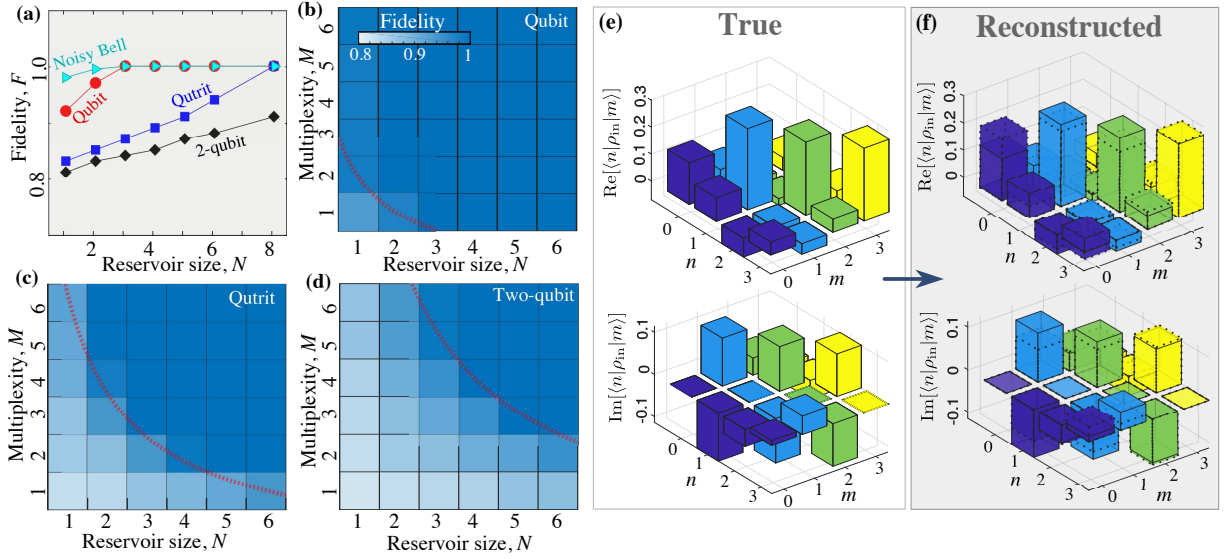}
\caption{The panel (a) shows the calculated fidelities as functions of reservoir size $N$ (number of fermions) for qubit, qutrit, two-qubit and noisy Bell states. Colour plots of fidelities calculated for (b) qubits, (c) qutrits and (d) two-qubits as functions of reservoir size $N$ and measurement multiplexity $M$ (the number of time instances when the occupation numbers are measured). The red dotted lines in each of the plots represent the relation $NM = D^2-1$. We find full tomographic reconstruction of the density matrices with fidelity $F=1$ for the regimes beyond the red dotted lines where $NM\ge D^2-1$. The data points are averaged over $10$ realizations of all random parameters. Panels (e) and (f) show the real and imaginary parts of an example two-qubit density matrix and the corresponding reconstruction by a QRST device with $N=2$, $M=6$ (dotted line) and $N=6$, $M=6$ (solid line) with the fidelities $0.9515\pm 10^{-5}$ and $1\pm10^{-5}$ respectively. The basis states $\ket{n}$ for $n=0,1,2$ and $3$ (same for $m$) represent the two-qubit basis states $\ket{00},~\ket{01},~\ket{10}$ and $\ket{11}$ respectively. Here we use $\tilde{J}/\gamma=1$, $\omega/\gamma=1$, $t_1 = 7.6\hbar/\gamma$, $\tau=1.5\hbar/\gamma$, $P/\gamma=0.3$, and the random parameters $\epsilon$ and $\varphi$ are chosen from the intervals $[0,~0.2]$ and $[0,2\pi]$ respectively.}
\label{FiniteTomographyError}
\end{figure*}

\textbf{Tomography in finite dimensions:--} Consider a system represented by a $D$ dimensional Hilbert space. The density matrices representing the quantum states of this system can be written as,
\begin{eqnarray}
\rho_\text{in} = \frac{1}{D}\left( \mathbbm{1}+ \sum_i s_i\zeta_i \right)
\label{FiniteDInput}
\end{eqnarray}
where $s_i$ are the $D^2-1$ independent parameters required to describe a state in $D$ dimensions, and $\zeta_i$ are the SU(D) generators that together with the identity matrix $\mathbbm{1}$ satisfy the completeness relation in the space of $D\times D$ matrices~\cite{Schlienz95}. The parameters $s_i$ are chosen using the Monte Carlo sampling technique with the constraint that all eigenvalues of $\rho_\text{in}$ are positive semidefinite (see Supplementary material). 
To perform tomography, we assign $ \textbf{Y}^\text{out} = \vec{\rho}_\text{in}$ where $\vec{\rho}_\text{in}$ represents the density matrix $\rho_\text{in}$ arranged in a column vector format. 
We find the output weight matrix $\mathcal{M}^\text{out}$ and the constant vector $\vec{m}$ by training the network with known examples of random density matrices. Here, training is equivalent to solving the matrix equation $\vec{\rho}_\text{in} =  \mathcal{M}^\text{out}\vec{n} + \vec{m}$. As the matrix equation corresponds to $D^2$ linear equations, the required minimum number of known examples of $\rho_\text{in}$ is $D^2$ for $D$ dimensional states. These training states require to be linearly independent, such that, each example represents an independent equation. In fact, we can consider $D^2$ number of randomly generated states for training, as it is statistically unlikely to randomly generate linearly dependent states. However, we train the network using the ridge regression technique~\cite{Lukosevicius12} with slightly larger number of examples than $D^2$ to avoid insufficient number of independent equations due to the rare generation of linearly dependent states. Once $ \mathcal{M}^\text{out}$ and $\vec{m}$ are determined, the density matrix of an input state is reconstructed from the output as $\vec{\rho}_\text{in}^\text{ tomo} = \textbf{Y}^\text{out} = \mathcal{M}^\text{out}\vec{n} + \vec{m}$, where $\vec{\rho}_\text{in}^\text{ tomo}$ is the vector form of the reconstructed density matrix $ \rho_\text{in}^\text{tomo}$. 
This reconstructed matrix $ \rho_\text{in}^\text{tomo}$ might be imprecise. To estimate possible errors in the tomography, we calculate the fidelity:
\begin{eqnarray}
F =  \left( \text{Tr}\left[ \sqrt{ \sqrt{\rho_\text{in}}\, \rho_\text{in}^\text{tomo}  \,\sqrt{\rho_\text{in}} } \right] \right)^2
\end{eqnarray}
In error-free tomography, $F=1$ and $F<1$ otherwise.

In Fig.~\ref{FiniteTomographyError}(a), we show the fidelity $F$ for different reservoir sizes $N$ and dimensions $D$. The fidelity systematically increases with increasing reservoir size for any $D$, and the fidelity reaches $1$ at $N=D^2-1$ for $D=2$ and $3$ respectively. We could not fully verify this relation for $D=4$ as it requires to simulate a quantum reservoir of size $N=15$, which is beyond our computational reach. However, we introduce the concept of "time multiplexing" in section~\ref{TimeMultiplexing} which drastically reduces the required size $N$.

\textbf{Typical state tomography:--} We have shown tomography of density matrices sampled in the full $D^2-1$ parameter space. However, quantum states generated in actual experiments are often restricted in a small subspace of the full parameter space. As an example for representing such states, we consider noisy Bell states: $\rho_\text{in} = (1-\epsilon)\ket{\psi}\bra{\psi} + (\epsilon/4) \mathbbm{1}$, where $| \psi \rangle = (\ket{00} + e^{-i\varphi}\ket{11})/\sqrt{2}$, and $\epsilon$ is a parameter quantifying the amount of noise. For training, we use  a supervised learning technique which requires example input states. We generate the example states $\rho_\text{in}$ with randomly chosen $\epsilon$ and $\varphi$ from the intervals $[0,0.2]$ and $[0,2\pi]$ respectively. Using the ridge regression method and the example states, we obtain the output weight matrix $\mathcal{M}^\text{out}$ and $\vec{m}$ (see Supplementary material). After the training, we use another set of randomly generated $\epsilon$ and $\varphi$ to obtain the corresponding $\rho_\text{in}$ for testing the reconstruction ability of QRST. While the full reconstruction of an arbitrary 2-qubit density matrix requires $15$ readout elements, full reconstruction of noisy Bell states requires only $2$ readout elements, see Fig.~\ref{FiniteTomographyError}(a). This reduction in required number of readout elements is due to the smaller number of independent parameters needed to represent the input states (training and testing are performed with the same class of states). Similar reasons manifest in the tomography of low rank density matrices~\cite{Gross10}, or in adaptive methods~\cite{Qi17}. However, QRST automatically learns this fact from training and applies it for tomography without any state-specific modification in the scheme.

\textbf{Time multiplexing:--} The simulations above show that a single measurement process (measuring average occupation numbers in a single physical setup) with a reservoir size $N=D^2-1$, that provides $D^2-1$ readout elements, is sufficient for full reconstruction of density matrices. However, note that one can increase the number of readout elements by performing measurements at multiple times. The input modes are coupled to the reservoir for a time interval $\tau$ between $t=t_1$ to $t=t_1+\tau$. One can measure the occupation numbers at $M$ different times $t=t_1+j\tau/M$ ($j=1,2,\dots M$) during this time interval. Now instead of $N$ readout elements in a reservoir size of $N$, one has $N\times M$ readout elements and thus the expected reservoir size $N = (D^2-1)/M$ to achieve a fidelity $F=1$. 
In Fig.~\ref{FiniteTomographyError} (b-d), we show the fidelity of the reconstructed density matrices as functions of the reservoir size $N$ and multiplexity $M$ that indeed confirms this scaling. In Fig. 2 (e) and (f), we show an example of a two-qubit density matrix and the corresponding reconstructions using ($N = 2$, $M=6$) and ($N = 6$, $M=6$).


\begin{figure}[]
\centering
\includegraphics[width=1\columnwidth]{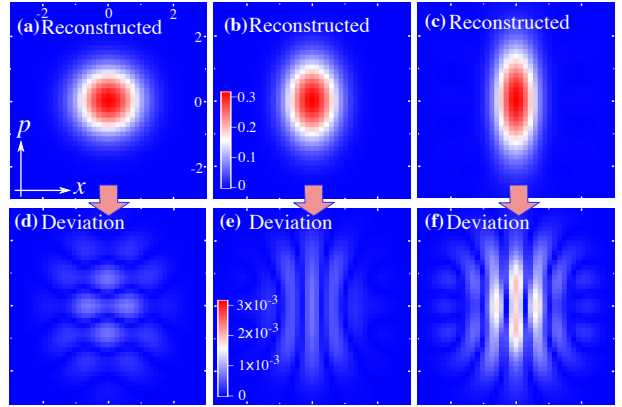}
\caption{Tomography of continuous variable quantum states. Here we show some considered examples of reconstructed squeezed-thermal states ((a) to (c)) and their deviations from the true states ((d) to (f)). The deviation is defined by $|W(\rho_\text{in}; x_i,p_j) - W^\text{tomo}(\rho_\text{in}; x_i,p_j)|$. Color scales for ((a) to (c)) and ((d) to (f)) are given in (b) and (e) respectively. We here use $16$ readout elements $(N=4,~M=4)$, $\tilde{J}/\gamma=1$, $\omega/\gamma=1$, $t_1 = 7.6\hbar/\gamma$, $\tau=1.5\hbar/\gamma$ and $P/\gamma=0.3$.}
\label{CVTomo}
\end{figure}
\begin{figure}[]
\centering
\includegraphics[width=1\columnwidth]{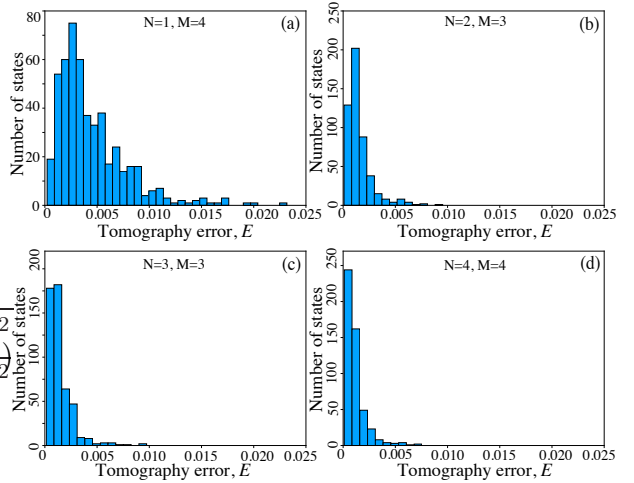}
\caption{Here we show the histograms for errors in estimating Wigner functions for increasing number of readout elements. Panels (a) to (d) show the histogram of $500$ squeezed-thermal states for their tomographic errors $E$ defined in Eq. (\ref{EQ_E}). The considered number of readout elements are $4$, $6$, $9$ and $16$ for panels (a) to (d) respectively. Here we use $\tilde{J}/\gamma=1$, $\omega/\gamma=1$, $t_1 = 7.6\hbar/\gamma$, $\tau=1.5\hbar/\gamma$ and $P/\gamma=0.3$.}
\label{CVTomoError}
\end{figure}

\textbf{Continuous variable tomography:--} We now consider continuous variable tomography using QRST. Despite an infinite (large) dimensionality of a continuous variable state, we show that an accurate tomography can be performed with a few readout elements in a single measurement process. For continuous variable states, it is convenient to represent them with Wigner functions $W(\rho_\text{in}; x_i,p_j)$, which are defined on a grid of continuous variables $x_i$ and $p_j$ for a given density matrix $\rho_\text{in}$.
We assign $Y_{ij}^\text{out} = W(\rho_\text{in}, x_i,p_j) $ as the output for QRST. Training provides the optimum output weight matrix and constants such that the reconstructed $W(\rho_\text{in}; x_i,p_j) = \sum_k \mathcal{M}_{ijk}^\text{out} n_k + m_{ij}$ has the minimum deviation from the corresponding known training Wigner functions. We use $96$ randomly generated squeezed-thermal states for training. In Fig.~\ref{CVTomo}, we show reconstructed Wigner functions of some randomly generated squeezed-thermal states (see Supplementary material) with a reservoir size $N=4$ and multiplexity $M=4$ and their deviations from the true Wigner functions. As a quantitative measure of the tomography error, we define:
\begin{eqnarray}
E = \sqrt{ \frac{ \sum_{i,j} [ W(\rho_\text{in}; x_i,p_j) - W^\text{tomo}(\rho_\text{in}; x_i,p_j) ]^2 }{ \sum_{i,j} [ W(\rho_\text{in}; x_i,p_j) + W^\text{tomo}(\rho_\text{in}; x_i,p_j) ]^2} }
\label{EQ_E}
\end{eqnarray}
where $W^\text{tomo}(\rho_\text{in}; x_i,p_j)$ is the reconstructed Wigner function corresponding to the true one $ W(\rho_\text{in}; x_i,p_j)$. In Fig.~\ref{CVTomoError}, we show the histogram of the estimated errors for different numbers of readout elements. We find the error systematically decreases with an increasing number of readout elements. In fact, the error is impressively low even with a small reservoir size like $N=4$ with $M=4$. Note that the effective dimension $D$ of the considered squeezed-thermal states is large. We identify the effective dimension $D$ as the smallest dimension for which the mean photon number becomes independent of $D$.  As shown in Fig.~\ref{CVTomo}, this small reservoir can reconstruct well the features of the Wigner functions. Already for a reservoir with $N=3$ and $M=3$, which is an even smaller reservoir with little multiplexity, the reconstruction errors are mostly close to zero. This demonstrates that QRST is a very powerful tool for quantum state tomography in the continuous variable domain.

\section{Theoretical generalisation}
Let us now explain the features observed in the simulations and prove their generalisation to arbitrary dimension. Since there are $D^2-1$ independent real parameters in a density matrix in order to estimate all of them the reservoir must have at least $N = D^2 - 1$ lattice sites (or $N=(D^2-1)/M$, if time-multiplexing is used).
In the numerics we find that for this number of sites QRST trained with $D^2$ random states reconstructs an \emph{arbitrary} input state perfectly, i.e. with unit fidelity.
This can be understood as follows.
Note that an arbitrary density matrix can be decomposed in terms of $D^2$ linearly independent states:
\begin{equation}
\rho_{\text{in}} = \sum_{i} \alpha_i \rho_i, \quad \textrm{with} \quad \sum_i \alpha_i = 1,
\label{EQ_ARBIN}
\end{equation}
where the coefficients $\alpha_i$ are not necessarily non-negative, but sum up to unity due to normalisation.
A random set of $D^2$ states is practically always linearly independent.
Therefore training of QRST with random states $\rho_i$ is equivalent to determining the matrix $\mathcal{M}^\text{out}$ and vector of constants $\vec m$ by solving the following set of $D^2$ linear equations:
\begin{equation}
\mathcal{M}^\text{out} \vec n_i + \vec m = \vec \rho_i,
\label{EQ_MTRAINED}
\end{equation}
where $\vec n_i$ is the vector of readout elements for random input state $\rho_i$ with vector representation $\vec \rho_i$.
In the Supplementary material we prove that the mean occupation number of a fermionic site of the reservoir 
can be represented as a positive operator valued measure (POVM) measurement on the input modes:
\begin{equation}
n_j = \langle \hat{c}^\dagger_j \hat{c}_j\rangle = \mathrm{tr}(\rho_\text{in} E_j),
\end{equation}
where $E_j$ is the POVM element corresponding to finding a fermion on the $j$th lattice site.
Hence the vector of readout elements corresponding to the arbitrary state in Eq. (\ref{EQ_ARBIN}) is given by $\vec n = \sum_i \alpha_i \vec n_i$.
It is now essential that our reconstruction procedure trained in Eq. (\ref{EQ_MTRAINED}) is a linear map on the readout vector $\vec n$. Indeed one readily verifies that $\mathcal{M}^\text{out} \vec n + \vec m = \vec \rho_{\text{in}}$, i.e., QRST perfectly recovers an arbitrary input state.

In conventional quantum state tomography, each von Neumann measurement basis provides at most $D-1$ independent real parameters characterising the input state. Accordingly one needs at least $D+1$ such measurement bases to reconstruct an arbitrary state in $D$ dimensional Hilbert space~\cite{WF89}. Typically more measurement bases are required due to complexity of these $D+1$ measurements as they necessarily require projections on entangled states~\cite{WPZ2011}. A way to reduce the number of measurement bases is to consider their generalisation in the form of informationally-complete POVMs. Typically these are also difficult to implement, see e.g.~\cite{REC2004,ZTE2010,KSE2012}, and it is our main point here that the informationally-complete set of $N$ POVMs we propose is practically implementable even for high dimensional input. Furthermore, an experimenter does not have to know the corresponding POVM elements. They are established indirectly via training and the output of QRST is the final density matrix.

\section{Error analysis}
As in any tomography scheme, the presence of measurement errors can introduce limitations. 
In a given run of the experiment, the measurement of the occupation number of a given node would give an integer number of particles. It is only by repeating the experiment multiple times that an average is obtained. The error in the measured occupation number is then given by:
\begin{eqnarray}
\sigma_{ \overline{n}_j }=\sigma_{n_j}/\sqrt{N_r}
\end{eqnarray}
where $N_r$ is the number of repetitions and $\sigma_{n_j} = \sqrt{ \expect{\hat{n}^2_j} - \expect{\hat{n}_j}^2} $ is the standard deviation in the particle number, which could be obtained from the density matrix. In addition to the quantum error, we also consider other random and systematic errors in the experimental setup.

\begin{figure}[]
\centering
\includegraphics[width=1\columnwidth]{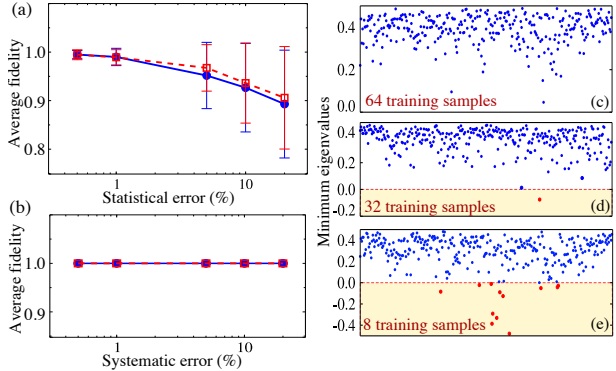}
\caption{We show the effects of statistical errors (a) and the systematic errors (b) in QRST returning states of a single qubit. We plot the average fidelity (averaged over $160$ random qubit states) for two reservoir sizes $N=3$ (blue solid line) and $N=4$ (red dotted line). The statistical and systematic error percentages are given by $ \sigma_r \times 100\%$ and $\sigma_s \times 100\%$ respectively. (c), (d) and (e) are showing the minimum eigenvalues of $320$ reconstructed density matrices for number of training samples $64$, $32$ and $8$ respectively. The positive and negative eigenvalues are indicated with blue and red points. Here we considered $20\%$ statistical and systematic errors both and reservoir of size $N=3$. We find that even with high errors in the output, the reconstructed density matrices show negative eigenvalues only rarely for small number of training samples.}
\label{MeasurementErrors}
\end{figure}

To characterize the possibility of further random errors, we consider that even after being repeated $N_r$ times, the evaluation of the occupation numbers with a given input state $\rho_\text{in}^{(i)}$ would still contain an error, where the actual value of the average occupation number is
\begin{eqnarray}
\tilde{n}_j = n_j(1+\sigma_r g_{r,i,j})
\end{eqnarray}
Here $g_{r,i,j}$ is a Gaussian random variable, which is different for different input states. $\sigma_r$ defines the overall size of the random errors and we allow it to be larger than the quantum error, potentially characterizing additional random errors in an experimental setup. In Fig.~\ref{MeasurementErrors}(a), we show how the average fidelity of single qubit states varies with the overall size of the random errors. Even with errors on the order of $10\%$, the fidelity remains above 90\%. Furthermore, the use of a larger reservoir allows for a slight compensation of errors and we can recall that random errors in an experimental setup could be further reduced by repeating the experiment more times.

Systematic errors in the average occupation numbers can be accounted for by defining:
\begin{eqnarray}
\tilde{n}_j = n_j(1+\sigma_s g_{s,j})
\end{eqnarray}
where $\sigma_s$ defines the size of the systematic errors, which are given by a Gaussian distribution and taken different for each node, independent of the input state. As shown in Fig.~\ref{MeasurementErrors}(b), the fidelity is unaffected by systematic errors. This is because, provided the systematic error is accounted for both in training and testing, the network would have learned to compensate the systematic errors in measurement. Mathematically, this is explained from the relation $\mathcal{M}^\text{out} ( \vec n_i + \delta  \vec  n)+ \vec m = \vec \rho_i$ where the systematic error $\delta n_j = \sigma_\text{s} g_\text{s,j} n_j$ is independent of the input state index $i$. Thus, we can redefine the vector $\vec m' = \vec m + \mathcal{M}^\text{out}   \delta  \vec  n$ to write $\mathcal{M}^\text{out} \vec n_i + \vec m' = \vec \rho_i$ where $ \vec m'$ remains a constant vector that does not depend on the input state and is determined by training.

The right panels in Fig.~\ref{MeasurementErrors} show the minimum eigenvalues of the reconstructed density matrices. We find that for sufficient training the eigenvalues of the reconstructed density matrix are nonnegative even when the output errors are as large as $20\%$. Although we have not explicitly imposed positivity on the reconstructed density matrix, the training process, which is performed only with physical density matrices, drives the neural network to form only positive density matrix at the output. However, practical limitations like an insufficient number of training samples, or extremely small reservoir size together with large measurement error can lead to the loss of positivity of the estimated density matrix, see Fig.~\ref{MeasurementErrors}(e). In such situations, a maximum likelihood estimation can be employed to find out the closest physical density matrix. A wealth of study has been devoted to explore this estimation~\cite{Banaszek99,Qi13,Shang17,Scholten18,Acharya19}. When needed, these methods can directly be applied as a final step in the QRST scheme. We emphasise again that our results strongly suggest that non-physical reconstructions do not arise for a sufficiently large number of training samples and note that the training has to be performed only once.

\section{Discussions}
\begin{figure}[]
\centering
\includegraphics[width=1\columnwidth]{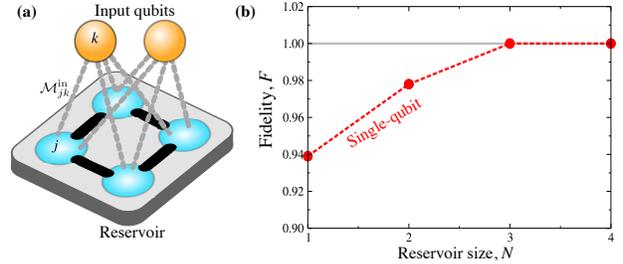}
\caption{Quantum state tomography with alternative form of coupling between the input mode and the reservoir sites. (a) shows the schematic diagram of the coupling between input qubits and the reservoir using quantum hopping. (b) shows the calculated fidelities as functions of reservoir size $N$ (with $M=1$) for single qubit states. Other parameters are taken same as mentioned in Fig.~\ref{FiniteTomographyError}.}
\label{JosephsonCoupling}
\end{figure}

For hardware implementation, our proposed platform can be realized in a wide range of systems in principle, e.g., arrays of semiconductor quantum dots, coupled superconducting qubits and trapped ions and atoms. Here we have considered a cascade formalism to model the coupling between the optical input state to the reservoir sites. Equivalently, for qubits realized in ions, atoms, spins and superconductors, the cascade coupling can be replaced with quantum hopping (Josephson type tunnelling) between the input qubits and the reservoir sites. This can be described by an interaction Hamiltonian: 
\begin{eqnarray}
\hat{H}_\text{int} = \sum_{jk}\mathcal{M}_{jk}^\text{in} ( \hat{c}_j^\dagger \hat{a}_k + \hat{a}_k^\dagger \hat{c}_j ), 
\end{eqnarray}
where the whole system is described by the total Hamiltonian $\hat{H}_\text{tot} = \hat{H} +\Theta(t-t_1)\hat{H}_\text{int}$. For spin systems, the same interaction Hamiltonian is applicable. In the language of spins $\hat{H}_\text{int}$ translates to $\hat{H}_\text{int} = \sum_{jk}\mathcal{M}_{jk}^\text{in} ( \hat{\sigma}_j^+  \hat{\sigma}_k^-  +  \hat{\sigma}_k^+  \hat{\sigma}_j^- )$ where the correspondence is made with $(\hat{c}_j,  \hat{c}_j^\dagger)\to (\hat{\sigma}_j^- , \hat{\sigma}_j^+)$ and $(\hat{a}_k, \hat{a}_k^\dagger ) \to (\hat{\sigma}_k^- ,\hat{\sigma}_k^+)$, and $\hat{\sigma}_m^{\pm} = (\hat{\sigma}_m^x \pm i \hat{\sigma}_m^y)/2 $ are the Pauli matrices for site $m$. This interaction is also known as the $XY$ interaction between the spins. In Fig.~\ref{JosephsonCoupling}, we show that QRST can also be performed by replacing the cascade coupling with $\hat{H}_\text{int} $. Exciton-polaritons in semiconducting microcavities (hybrid light-matter quasiparticles) are also suitable for realizing quantum reservoirs as they naturally couple with external optical fields and can reach strongly interacting regimes. Here the coupling between the input state and the reservoir is realisable either with the cascade method or with the quantum hopping. Moreover, while implementing a quantum reservoir in a physical system our protocol does not require precise control on the system parameters, rather, randomness is a useful resource for successful quantum state tomography.

Without time multiplexing, the required reservoir size for $D$ dimensional tomography is $N=D^2-1$. With the time multiplexing ($M$) the required size reduces to $N = (D^2-1)/M$. An alternative route to reduce $N$ would be measuring the quantum correlations between the nodes. Since the number of possible quantum correlations increases exponentially with $N$, the required reservoir size would decreases to the order of $\log(D)$. However, we think that measuring exponentially large number of quantum correlations would be a much challenging task, compared to measuring the average occupation numbers.

\section{Conclusion}
We have presented quantum reservoir state tomography as a platform for quantum state reconstruction. Unlike existing schemes for specific quantum states either in finite-dimensional or continuous-variable domain, our scheme operates the same for any quantum state. Within our scheme, tomography is accomplished with a single measurement process of intensity, which does not require photon number resolution or correlated detection, in contrast to previous tomography schemes that use multiple measurements in different bases and correlation measurements of complex informationally-complete generalised measurement schemes. Here the challenge can be in achieving a reservoir that can be large for higher dimensional tomography. However, we have shown that the required size of a quantum reservoir can be drastically reduced by measuring occupation numbers at the reservoir lattice sites at multiple times (time multiplexing) for readout. It is thus an experimentally friendly, practically scalable scheme that can be universally applied to any quantum state.

\section*{Acknowledgement} 
This work is supported by the Singapore Ministry of Education Academic Research Fund Tier 2, Project No. MOE2015-T2-2-034, MOE2017-T2-1-001, and MOE2019-T2-1-004. MM and AO acknowledge support from the National Science Center, Poland grant No. 2016/22/E/ST3/00045. The authors thank Tanjung Krisnanda for helpful discussions during the preparation of this manuscript.


\begin{thebibliography}{10}
\providecommand{\url}[1]{#1}
\csname url@samestyle\endcsname
\providecommand{\newblock}{\relax}
\providecommand{\bibinfo}[2]{#2}
\providecommand{\BIBentrySTDinterwordspacing}{\spaceskip=0pt\relax}
\providecommand{\BIBentryALTinterwordstretchfactor}{4}
\providecommand{\BIBentryALTinterwordspacing}{\spaceskip=\fontdimen2\font plus
\BIBentryALTinterwordstretchfactor\fontdimen3\font minus
  \fontdimen4\font\relax}
\providecommand{\BIBforeignlanguage}[2]{{%
\expandafter\ifx\csname l@#1\endcsname\relax
\typeout{** WARNING: IEEEtran.bst: No hyphenation pattern has been}%
\typeout{** loaded for the language `#1'. Using the pattern for}%
\typeout{** the default language instead.}%
\else
\language=\csname l@#1\endcsname
\fi
#2}}
\providecommand{\BIBdecl}{\relax}
\BIBdecl

\bibitem{Biamonte17}
\BIBentryALTinterwordspacing
J.~Biamonte, P.~Wittek, N.~Pancotti, P.~Rebentrost, N.~Wiebe, and S.~Lloyd,
  ``Quantum machine learning,'' \emph{Nature}, vol. 549, p. 195, 09 2017.
  [Online]. Available: \url{https://doi.org/10.1038/nature23474}
\BIBentrySTDinterwordspacing

\bibitem{Dunjko16}
\BIBentryALTinterwordspacing
V.~Dunjko, J.~M. Taylor, and H.~J. Briegel, ``Quantum-enhanced machine
  learning,'' \emph{Physical Review Letters}, vol. 117, no.~13, pp. 130\,501--,
  09 2016. [Online]. Available:
  \url{https://link.aps.org/doi/10.1103/PhysRevLett.117.130501}
\BIBentrySTDinterwordspacing

\bibitem{Paparo14}
\BIBentryALTinterwordspacing
G.~D. Paparo, V.~Dunjko, A.~Makmal, M.~A. Martin-Delgado, and H.~J. Briegel,
  ``Quantum speedup for active learning agents,'' \emph{Physical Review X},
  vol.~4, no.~3, pp. 031\,002--, 07 2014. [Online]. Available:
  \url{https://link.aps.org/doi/10.1103/PhysRevX.4.031002}
\BIBentrySTDinterwordspacing

\bibitem{Neigovzen09}
\BIBentryALTinterwordspacing
R.~Neigovzen, J.~L. Neves, R.~Sollacher, and S.~J. Glaser, ``Quantum pattern
  recognition with liquid-state nuclear magnetic resonance,'' \emph{Physical
  Review A}, vol.~79, no.~4, pp. 042\,321--, 04 2009. [Online]. Available:
  \url{https://link.aps.org/doi/10.1103/PhysRevA.79.042321}
\BIBentrySTDinterwordspacing

\bibitem{Benedetti16}
\BIBentryALTinterwordspacing
M.~Benedetti, J.~Realpe-G{\'o}mez, R.~Biswas, and A.~Perdomo-Ortiz,
  ``Estimation of effective temperatures in quantum annealers for sampling
  applications: A case study with possible applications in deep learning,''
  \emph{Physical Review A}, vol.~94, no.~2, pp. 022\,308--, 08 2016. [Online].
  Available: \url{https://link.aps.org/doi/10.1103/PhysRevA.94.022308}
\BIBentrySTDinterwordspacing

\bibitem{Alvarez-Rodriguez17}
\BIBentryALTinterwordspacing
U.~Alvarez-Rodriguez, L.~Lamata, P.~Escandell-Montero, J.~Mart{\'\i}n-Guerrero,
  and E.~Solano, ``Supervised quantum learning without measurements,''
  \emph{Scientific Reports}, vol.~7, no.~1, p. 13645, 2017. [Online].
  Available: \url{https://doi.org/10.1038/s41598-017-13378-0}
\BIBentrySTDinterwordspacing

\bibitem{Fujii17}
\BIBentryALTinterwordspacing
K.~Fujii and K.~Nakajima, ``Harnessing disordered-ensemble quantum dynamics for
  machine learning,'' \emph{Physical Review Applied}, vol.~8, no.~2, pp.
  024\,030--, 08 2017. [Online]. Available:
  \url{https://link.aps.org/doi/10.1103/PhysRevApplied.8.024030}
\BIBentrySTDinterwordspacing

\bibitem{Farhi18}
\BIBentryALTinterwordspacing
E.~Farhi and H.~Neven, ``Classification with quantum neural networks on near
  term processors,'' 2018. [Online]. Available:
  \url{https://ui.adsabs.harvard.edu/\#abs/2018arXiv180206002F}
\BIBentrySTDinterwordspacing

\bibitem{Grant18}
\BIBentryALTinterwordspacing
E.~Grant, M.~Benedetti, S.~Cao, A.~Hallam, J.~Lockhart, V.~Stojevic, A.~G.
  Green, and S.~Severini, ``Hierarchical quantum classifiers,'' \emph{npj
  Quantum Information}, vol.~4, no.~1, p.~65, 2018. [Online]. Available:
  \url{https://doi.org/10.1038/s41534-018-0116-9}
\BIBentrySTDinterwordspacing

\bibitem{Amin18}
\BIBentryALTinterwordspacing
M.~H. Amin, E.~Andriyash, J.~Rolfe, B.~Kulchytskyy, and R.~Melko, ``Quantum
  boltzmann machine,'' \emph{Physical Review X}, vol.~8, no.~2, pp. 021\,050--,
  05 2018. [Online]. Available:
  \url{https://link.aps.org/doi/10.1103/PhysRevX.8.021050}
\BIBentrySTDinterwordspacing

\bibitem{Tanaka19}
G.~Tanaka, T.~Yamane, J.~B. H{\'e}roux, R.~Nakane, N.~Kanazawa, S.~Takeda,
  H.~Numata, D.~Nakano, and A.~Hirose, ``Recent advances in physical reservoir
  computing: A review,'' \emph{Neural Networks}, vol. 115, pp. 100--123, 2019.

\bibitem{Lukosevicius12}
M.~Luko{\v s}evi{\v c}ius, \emph{A Practical Guide to Applying Echo State
  Networks}, G.~Montavon, G.~B. Orr, and K.-R. M{\"u}ller, Eds.\hskip 1em plus
  0.5em minus 0.4em\relax Berlin, Heidelberg: Springer Berlin Heidelberg, 2012.

\bibitem{Bengio94}
Y.~Bengio, P.~Simard, and P.~Frasconi, ``Learning long-term dependencies with
  gradient descent is difficult,'' \emph{IEEE Transactions on Neural Networks},
  vol.~5, no.~2, pp. 157--166, 1994.

\bibitem{Ghosh19}
\BIBentryALTinterwordspacing
S.~Ghosh, A.~Opala, M.~Matuszewski, T.~Paterek, and T.~C.~H. Liew, ``Quantum
  reservoir processing,'' \emph{npj Quantum Information}, vol.~5, no.~1, p.~35,
  2019. [Online]. Available: \url{https://doi.org/10.1038/s41534-019-0149-8}
\BIBentrySTDinterwordspacing

\bibitem{Shadbolt11}
\BIBentryALTinterwordspacing
P.~J. Shadbolt, M.~R. Verde, A.~Peruzzo, A.~Politi, A.~Laing, M.~Lobino,
  J.~C.~F. Matthews, M.~G. Thompson, and J.~L. O'Brien, ``Generating,
  manipulating and measuring entanglement and mixture with a reconfigurable
  photonic circuit,'' \emph{Nature Photonics}, vol.~6, p.~45, 12 2011.
  [Online]. Available: \url{https://doi.org/10.1038/nphoton.2011.283}
\BIBentrySTDinterwordspacing

\bibitem{Lvovsky09}
\BIBentryALTinterwordspacing
A.~I. Lvovsky and M.~G. Raymer, ``Continuous-variable optical quantum-state
  tomography,'' \emph{Reviews of Modern Physics}, vol.~81, no.~1, pp. 299--332,
  03 2009. [Online]. Available:
  \url{https://link.aps.org/doi/10.1103/RevModPhys.81.299}
\BIBentrySTDinterwordspacing

\bibitem{James01}
\BIBentryALTinterwordspacing
D.~F.~V. James, P.~G. Kwiat, W.~J. Munro, and A.~G. White, ``Measurement of
  qubits,'' \emph{Physical Review A}, vol.~64, no.~5, pp. 052\,312--, 10 2001.
  [Online]. Available:
  \url{https://link.aps.org/doi/10.1103/PhysRevA.64.052312}
\BIBentrySTDinterwordspacing

\bibitem{Haffner05}
\BIBentryALTinterwordspacing
H.~H{\"a}ffner, W.~H{\"a}nsel, C.~F. Roos, J.~Benhelm, D.~Chek-al kar,
  M.~Chwalla, T.~K{\"o}rber, U.~D. Rapol, M.~Riebe, P.~O. Schmidt, C.~Becher,
  O.~G{\"u}hne, W.~D{\"u}r, and R.~Blatt, ``Scalable multiparticle entanglement
  of trapped ions,'' \emph{Nature}, vol. 438, p. 643, 12 2005. [Online].
  Available: \url{https://doi.org/10.1038/nature04279}
\BIBentrySTDinterwordspacing

\bibitem{Lu07}
\BIBentryALTinterwordspacing
C.-Y. Lu, X.-Q. Zhou, O.~G{\"u}hne, W.-B. Gao, J.~Zhang, Z.-S. Yuan, A.~Goebel,
  T.~Yang, and J.-W. Pan, ``Experimental entanglement of six photons in graph
  states,'' \emph{Nature Physics}, vol.~3, p.~91, 01 2007. [Online]. Available:
  \url{https://doi.org/10.1038/nphys507}
\BIBentrySTDinterwordspacing

\bibitem{Ferrie14}
\BIBentryALTinterwordspacing
C.~Ferrie, ``Self-guided quantum tomography,'' \emph{Physical Review Letters},
  vol. 113, no.~19, pp. 190\,404--, 11 2014. [Online]. Available:
  \url{https://link.aps.org/doi/10.1103/PhysRevLett.113.190404}
\BIBentrySTDinterwordspacing

\bibitem{Chapman16}
\BIBentryALTinterwordspacing
R.~J. Chapman, C.~Ferrie, and A.~Peruzzo, ``Experimental demonstration of
  self-guided quantum tomography,'' \emph{Physical Review Letters}, vol. 117,
  no.~4, pp. 040\,402--, 07 2016. [Online]. Available:
  \url{https://link.aps.org/doi/10.1103/PhysRevLett.117.040402}
\BIBentrySTDinterwordspacing

\bibitem{Qi17}
\BIBentryALTinterwordspacing
B.~Qi, Z.~Hou, Y.~Wang, D.~Dong, H.-S. Zhong, L.~Li, G.-Y. Xiang, H.~M.
  Wiseman, C.-F. Li, and G.-C. Guo, ``Adaptive quantum state tomography via
  linear regression estimation: Theory and two-qubit experiment,'' \emph{npj
  Quantum Information}, vol.~3, no.~1, p.~19, 2017. [Online]. Available:
  \url{https://doi.org/10.1038/s41534-017-0016-4}
\BIBentrySTDinterwordspacing

\bibitem{Torlai18}
\BIBentryALTinterwordspacing
G.~Torlai, G.~Mazzola, J.~Carrasquilla, M.~Troyer, R.~Melko, and G.~Carleo,
  ``Neural-network quantum state tomography,'' \emph{Nature Physics}, vol.~14,
  no.~5, pp. 447--450, 2018. [Online]. Available:
  \url{https://doi.org/10.1038/s41567-018-0048-5}
\BIBentrySTDinterwordspacing

\bibitem{Quek18}
Y.~{Quek}, S.~{Fort}, and H.~{Khoon Ng}, ``{Adaptive Quantum State Tomography
  with Neural Networks},'' \emph{arXiv e-prints}, p. arXiv:1812.06693, Dec
  2018.

\bibitem{Carrasquilla19}
\BIBentryALTinterwordspacing
J.~Carrasquilla, G.~Torlai, R.~G. Melko, and L.~Aolita, ``Reconstructing
  quantum states with generative models,'' \emph{Nature Machine Intelligence},
  vol.~1, no.~3, pp. 155--161, 2019. [Online]. Available:
  \url{https://doi.org/10.1038/s42256-019-0028-1}
\BIBentrySTDinterwordspacing

\bibitem{Gross10}
\BIBentryALTinterwordspacing
D.~Gross, Y.-K. Liu, S.~T. Flammia, S.~Becker, and J.~Eisert, ``Quantum state
  tomography via compressed sensing,'' \emph{Physical Review Letters}, vol.
  105, no.~15, pp. 150\,401--, 10 2010. [Online]. Available:
  \url{https://link.aps.org/doi/10.1103/PhysRevLett.105.150401}
\BIBentrySTDinterwordspacing

\bibitem{Cramer10}
\BIBentryALTinterwordspacing
M.~Cramer, M.~B. Plenio, S.~T. Flammia, R.~Somma, D.~Gross, S.~D. Bartlett,
  O.~Landon-Cardinal, D.~Poulin, and Y.-K. Liu, ``Efficient quantum state
  tomography,'' \emph{Nature Communications}, vol.~1, p. 149, 12 2010.
  [Online]. Available: \url{https://doi.org/10.1038/ncomms1147}
\BIBentrySTDinterwordspacing

\bibitem{Oren17}
\BIBentryALTinterwordspacing
D.~Oren, M.~Mutzafi, Y.~C. Eldar, and M.~Segev, ``Quantum state tomography with
  a single measurement setup,'' \emph{Optica}, vol.~4, no.~8, pp. 993--999,
  2017. [Online]. Available:
  \url{http://www.osapublishing.org/optica/abstract.cfm?URI=optica-4-8-993}
\BIBentrySTDinterwordspacing

\bibitem{Titchener18}
\BIBentryALTinterwordspacing
J.~G. Titchener, M.~Gr{\"a}fe, R.~Heilmann, A.~S. Solntsev, A.~Szameit, and
  A.~A. Sukhorukov, ``Scalable on-chip quantum state tomography,'' \emph{npj
  Quantum Information}, vol.~4, no.~1, p.~19, 2018. [Online]. Available:
  \url{https://doi.org/10.1038/s41534-018-0063-5}
\BIBentrySTDinterwordspacing

\bibitem{Hofheinz09}
\BIBentryALTinterwordspacing
M.~Hofheinz, H.~Wang, M.~Ansmann, R.~C. Bialczak, E.~Lucero, M.~Neeley, A.~D.
  O'Connell, D.~Sank, J.~Wenner, J.~M. Martinis, and A.~N. Cleland,
  ``Synthesizing arbitrary quantum states in a superconducting resonator,''
  \emph{Nature}, vol. 459, p. 546, 05 2009. [Online]. Available:
  \url{https://doi.org/10.1038/nature08005}
\BIBentrySTDinterwordspacing

\bibitem{Landon-Cardinal18}
\BIBentryALTinterwordspacing
O.~Landon-Cardinal, L.~C.~G. Govia, and A.~A. Clerk, ``Quantitative tomography
  for continuous variable quantum systems,'' \emph{Physical Review Letters},
  vol. 120, no.~9, pp. 090\,501--, 03 2018. [Online]. Available:
  \url{https://link.aps.org/doi/10.1103/PhysRevLett.120.090501}
\BIBentrySTDinterwordspacing

\bibitem{Carmichael93}
\BIBentryALTinterwordspacing
H.~J. Carmichael, ``Quantum trajectory theory for cascaded open systems,''
  \emph{Physical Review Letters}, vol.~70, no.~15, pp. 2273--2276, 04 1993.
  [Online]. Available:
  \url{https://link.aps.org/doi/10.1103/PhysRevLett.70.2273}
\BIBentrySTDinterwordspacing

\bibitem{Gardiner93}
\BIBentryALTinterwordspacing
C.~W. Gardiner, ``Driving a quantum system with the output field from another
  driven quantum system,'' \emph{Physical Review Letters}, vol.~70, no.~15, pp.
  2269--2272, 04 1993. [Online]. Available:
  \url{https://link.aps.org/doi/10.1103/PhysRevLett.70.2269}
\BIBentrySTDinterwordspacing

\bibitem{Schlienz95}
\BIBentryALTinterwordspacing
J.~Schlienz and G.~Mahler, ``Description of entanglement,'' \emph{Physical
  Review A}, vol.~52, no.~6, pp. 4396--4404, 12 1995. [Online]. Available:
  \url{https://link.aps.org/doi/10.1103/PhysRevA.52.4396}
\BIBentrySTDinterwordspacing

\bibitem{WF89}
W.~K. Wootters and B.~D. Fields, ``Optimal state-determination by mutually
  unbiased measurements,'' \emph{Annals of Physics}, vol. 191, pp. 363--381,
  1989.

\bibitem{WPZ2011}
M.~Wie\'sniak, T.~Paterek, and A.~Zeilinger, ``Entanglement in mutually
  unbiased bases,'' \emph{New Journal of Physics}, vol.~13, p. 053047, 2011.

\bibitem{REC2004}
J.~Rehacek, B.-G. Englert, and D.~Kaszlikowski, ``Minimal qubit tomography,''
  \emph{Physical Review A}, vol.~70, p. 052321, 2004.

\bibitem{ZTE2010}
H.~Zhu, Y.~S. Teo, and B.-G. Englert, ``Two-qubit symmetric informationally
  complete positive-operator-valued measures,'' \emph{Physical Review A},
  vol.~82, p. 042308, 2010.

\bibitem{KSE2012}
A.~Kalev, J.~Shang, and B.-G. Englert, ``Experimental proposal for symmetric
  minimal two-qubit state tomography,'' \emph{Physical Review A}, vol.~85, p.
  052115, 2012.

\bibitem{Banaszek99}
\BIBentryALTinterwordspacing
K.~Banaszek, G.~M. D'Ariano, M.~G.~A. Paris, and M.~F. Sacchi,
  ``Maximum-likelihood estimation of the density matrix,'' \emph{Physical
  Review A}, vol.~61, no.~1, pp. 010\,304--, 12 1999. [Online]. Available:
  \url{https://link.aps.org/doi/10.1103/PhysRevA.61.010304}
\BIBentrySTDinterwordspacing

\bibitem{Qi13}
\BIBentryALTinterwordspacing
B.~Qi, Z.~Hou, L.~Li, D.~Dong, G.~Xiang, and G.~Guo, ``Quantum state tomography
  via linear regression estimation,'' \emph{Scientific Reports}, vol.~3, no.~1,
  p. 3496, 2013. [Online]. Available: \url{https://doi.org/10.1038/srep03496}
\BIBentrySTDinterwordspacing

\bibitem{Shang17}
\BIBentryALTinterwordspacing
J.~Shang, Z.~Zhang, and H.~K. Ng, ``Superfast maximum-likelihood reconstruction
  for quantum tomography,'' \emph{Physical Review A}, vol.~95, no.~6, pp.
  062\,336--, 06 2017. [Online]. Available:
  \url{https://link.aps.org/doi/10.1103/PhysRevA.95.062336}
\BIBentrySTDinterwordspacing

\bibitem{Scholten18}
\BIBentryALTinterwordspacing
T.~L. Scholten and R.~Blume-Kohout, ``Behavior of the maximum likelihood in
  quantum state tomography,'' vol.~20, no.~2, p. 023050, 2018. [Online].
  Available: \url{http://dx.doi.org/10.1088/1367-2630/aaa7e2}
\BIBentrySTDinterwordspacing

\bibitem{Acharya19}
\BIBentryALTinterwordspacing
A.~Acharya, T.~Kypraios, and M.~Gu{\c t}{\u a}, ``A comparative study of
  estimation methods in quantum tomography,'' vol.~52, no.~23, p. 234001, 2019.
  [Online]. Available: \url{http://dx.doi.org/10.1088/1751-8121/ab1958}
\BIBentrySTDinterwordspacing

\end{thebibliography}

\end{document}